\documentclass[11pt]{article}
\font\cero=cmbx10 scaled 1728 %
\font\uno=cmcsc10 scaled 1200 %
\font\dos=cmti10 scaled 1200 %
\font\tres=cmbx12 scaled 1200 %
\title{\cero Transfer matrices for piecewise constant potentials}
\author{{\uno G.F.\ Torres del Castillo} \\
{\dos Departamento de F\'{\i}sica Matem\'atica, Instituto de Ciencias} \\
{\dos Universidad Aut\'onoma de Puebla, 72570 Puebla, Pue., M\'exico} \\
{\uno I. Rubalcava Garc\'{\i}a} \\
{\dos Facultad de Ciencias F\'{\i}sico Matem\'aticas} \\
{\dos Universidad Aut\'onoma de Puebla, Apartado postal 1152,} \\
{\dos 72001 Puebla, Pue., M\'exico}}
\date{ }
\begin{document}
\maketitle
\section*{ }
By expressing the time-independent Schr\"odinger equation in one
dimension as a system of two first-order differential equations, the
transfer matrix for a rectangular potential barrier is obtained
making use of the matrix exponential. It is shown that the transfer
matrix allows one to find the bound states and the quasinormal
modes. A similar treatment for the one-dimensional propagation of
electromagnetic waves in a homogeneous medium is
also presented.\\[1ex]
{\it Keywords:} Scattering; transfer matrix; quasinormal modes;
layered systems \\[2ex]
Expresando la ecuaci\'on de Schr\"odinger independiente del tiempo
en una dimensi\'on como un sistema de dos ecuaciones diferenciales
de primer orden, se obtiene la matriz de transferencia para una
barrera de potencial rectangular haciendo uso de la exponencial de
matrices. Se muestra que la matriz de transferencia permite hallar
los estados ligados y los modos cuasinormales. Se presenta tambi\'en
un tratamiento similar para la propagaci\'on unidimensional de ondas
electromagn\'eticas en un medio
homog\'eneo.\\[1ex]
{\it Descriptores:} Dispersi\'on; matriz de transferencia; modos
cuasinormales; sistemas en capas \\[2ex]
PACS: 03.65.-w, 02.10.Ud

\section*{\tres 1. Introduction}
A standard problem in elementary quantum mechanics is that of
finding the reflection and transmission amplitudes for the
scattering produced by a potential barrier, or well, in one
dimension (see, {\em e.g.}, Refs.\ 1--4). The reflection and
transmission amplitudes are conveniently arranged in the transfer
matrix, which relates the wave function at both sides of the
potential barrier, in such a way that the effect of two or more
potential barriers is readily obtained by means of the product of
the corresponding transfer matrices (see, {\em e.g.}, Ref.\ 5 and
the references cited therein). A similar result applies for the
one-dimensional propagation of electromagnetic waves in layered
media (see, {\em e.g.}, Ref.\ 6). In fact, the transfer matrices can
be defined in all cases where there is an output that depends
linearly on an input; some important examples, apart from the two
already mentioned, are the electric circuits and optical systems. In
the cases considered here, the transfer matrices are $2 \times 2$
complex matrices but, depending on the equations involved (more
specifically, the number of variables and the differential order),
the size of the transfer matrices do vary.

The aim of this paper is to show that the transfer matrix for a
rectangular potential barrier (and, therefore, for a piecewise
constant potential) can be easily obtained integrating the
time-independent Schr\"odinger equation in one dimension by means of
the matrix exponential. The time-independent Schr\"odinger equation
in one dimension, being a second-order ordinary differential
equation, is equivalent to a system of two coupled first-order
differential equations and, only in the case of a (piecewise)
constant potential, this system can be easily integrated using the
matrix exponential. We also show that making use of the transfer
matrix one can find the bound states and the quasinormal modes. The
transfer matrix for the one-dimensional propagation of
electromagnetic waves in a medium with a piecewise constant
refractive index is obtained in a similar manner, without employing
the Fresnel coefficients.

In Sec.\ 2 an elementary discussion about the transfer matrices for
the one-dimensional Schr\"odinger equation is given (see also Ref.\
5 and the references cited therein). In Sec.\ 3 the transfer matrix
for a rectangular barrier is obtained making use of the matrix
exponential; the bound states and quasinormal modes are then found
starting from the transfer matrix. In Sec.\ 4 a similar derivation
for the case of the one-dimensional propagation of electromagnetic
waves in layered media is given.

\section*{\tres 2. Transfer matrices}
The solutions of the time-independent Schr\"odinger equation
\begin{equation}
- \frac{\hbar^{2}}{2m} \frac{{\rm d}^{2} \psi}{{\rm d}x^{2}} + V(x)
\psi = E \psi \label{sch}
\end{equation}
with a given short-range potential $V(x)$, which vanishes outside
the interval $a \leq x \leq b$, can be expressed in the form
\begin{equation}
\psi (x) = \left\{ \begin{array}{ll} A_{1} {\rm e}^{{\rm i} k(x-a)}
+ A_{2} {\rm e}^{- {\rm i} k(x-a)}, & {\rm for\ } x < a,
\\
B_{1} {\rm e}^{{\rm i} k(x-b)} + B_{2} {\rm e}^{- {\rm i} k(x-b)},
& {\rm for\ } x > b, \\
u(x), & {\rm for\ } a \leq x \leq b, \end{array}\right. \label{sol}
\end{equation}
where $k \equiv \sqrt{2mE}/\hbar$, $A_{1}$, $A_{2}$, $B_{1}$,
$B_{2}$ are constants and $u(x)$ is a function that depends on the
explicit form of the potential $V(x)$. By imposing the usual
conditions of continuity of $\psi(x)$ and its derivative at $x = a$
and $x = b$, a linear relation of the form
\begin{equation}
\left( \begin{array}{c} A_{1} \\ A_{2} \end{array} \right) = M
\left( \begin{array}{c} B_{1} \\ B_{2} \end{array} \right)
\label{tm}
\end{equation}
can be obtained, where $M$ is some $2 \times 2$ complex matrix (the
transfer matrix), which depends on $V(x)$ and the value of $k$.

Assuming that $V(x)$ is real, Eq.\ (\ref{sch}) implies that the
probability current density
\[
j(x) = \frac{\hbar}{2 {\rm i} m} \left( \psi^{*} \frac{{\rm d}
\psi}{{\rm d}x} - \psi \frac{{\rm d} \psi^{*}}{{\rm d}x} \right),
\]
where ${}^{*}$ denotes complex conjugation, satisfies the continuity
equation, ${\rm d}j/{\rm d}x = 0$, that is, $j(x) = {\rm const.}$;
then, making use of Eq.\ (\ref{sol}), one finds that, for $k$ real
\begin{equation}
|A_{1}|^{2} - |A_{2}|^{2} = |B_{1}|^{2} - |B_{2}|^{2}. \label{uni}
\end{equation}
Using the fact that
\[
|A_{1}|^{2} - |A_{2}|^{2} = \left( \begin{array}{c} A_{1} \\ A_{2}
\end{array} \right)^{\dag} \left( \begin{array}{rr} 1 & 0 \\ 0 & -1
\end{array} \right) \left( \begin{array}{c} A_{1} \\ A_{2}
\end{array} \right),
\]
where the ${}^{\dag}$ denotes the Hermitian adjoint, and Eq.\
(\ref{tm}) one finds that Eq.\ (\ref{uni}) is equivalent to
\begin{equation}
M^{\dag} \left( \begin{array}{rr} 1 & 0 \\ 0 & -1
\end{array} \right) M = \left( \begin{array}{rr} 1 & 0 \\ 0 & -1
\end{array} \right). \label{suni}
\end{equation}
The complex $2 \times 2$ matrices satisfying Eq.\ (\ref{suni}) form
a group with the usual matrix multiplication (see below). Equation
(\ref{suni}) implies that the modulus of $\det M$ is equal to 1.

The entries of the transfer matrix are related to the reflection and
transmission amplitudes of the potential $V(x)$, denoted by $r$ and
$t$, respectively. When there are no waves coming from the right
($B_{2} = 0$), there exist solutions of the Schr\"odinger equation
of the form
\begin{equation}
\psi (x) = \left\{ \begin{array}{ll} {\rm e}^{{\rm i} k(x-a)} + r
{\rm e}^{- {\rm i} k(x-a)}, & {\rm for\ } x < a, \\
t {\rm e}^{{\rm i} k(x-b)}, & {\rm for\ } x > b, \\
u_{1}(x), & {\rm for\ } a \leq x \leq b, \end{array}\right.
\label{sol1}
\end{equation}
that is, solutions of the form (\ref{sol}) with $A_{1} = 1$, $A_{2}
= r$, and $B_{1} = t$. Thus, from Eq.\ (\ref{tm}) it follows that
\[
M = \left( \begin{array}{cc} 1/t & M_{12} \\ r/t & M_{22}
\end{array} \right),
\]
with $M_{12}$ and $M_{22}$ not yet identified, and from Eq.\
(\ref{uni}) we obtain the well-known relation
\begin{equation}
1 - |r|^{2} = |t|^{2}. \label{cons}
\end{equation}
(In most textbooks the reflection and transmission amplitudes are
defined by means of expressions similar to Eq.\ (\ref{sol1}), with
$r$ and $t$ being the coefficients of ${\rm e}^{- {\rm i} kx}$ and
${\rm e}^{{\rm i} kx}$ and therefore, the amplitudes $r$ and $t$
defined by Eq.\ (\ref{sol1}) differ from those usually employed by
factors ${\rm e}^{{\rm i}ka}$ and ${\rm e}^{-{\rm i}kb}$,
respectively.)

Since $V(x)$ is real, for $k$ real the complex conjugate of the
solution (\ref{sol1})
\begin{equation}
\psi^{*} (x) = \left\{ \begin{array}{ll} r^{*} {\rm e}^{{\rm i}
k(x-a)} + {\rm e}^{- {\rm i} k(x-a)}, & {\rm for\ } x < a, \\
t^{*} {\rm e}^{- {\rm i} k(x-b)}, & {\rm for\ } x > b, \\
u^{*}_{1}(x), & {\rm for\ } a \leq x \leq b, \end{array}\right.
\label{sol2}
\end{equation}
is also a solution of the Schr\"odinger equation. Substituting the
coefficients appearing in Eq.\ (\ref{sol2}) into Eq.\ (\ref{tm}) we
find that
\begin{equation}
M = \left( \begin{array}{cc} 1/t & r^{*}/t^{*} \\ r/t & 1/t^{*}
\end{array} \right). \label{su}
\end{equation}
Then, according to Eq.\ (\ref{cons}), $\det M = 1$, which means that
(for $k$ real) the transfer matrix belongs to the group SU(1,1),
formed by the $2 \times 2$ complex matrices with unit determinant
that satisfy Eq.\ (\ref{suni}).

The reflection and transmission amplitudes of the potential $V(x)$
for waves incident from the right, $r'$ and $t'$, respectively, need
not coincide with $r$ and $t$. In fact, from Eqs.\ (\ref{tm}) and
(\ref{su}), setting $A_{1} = 0$ and $B_{2} = 1$, we must have
\[
\left( \begin{array}{c} 0 \\ t' \end{array} \right) = \left(
\begin{array}{cc} 1/t & r^{*}/t^{*} \\ r/t & 1/t^{*} \end{array}
\right) \left( \begin{array}{c} r' \\ 1 \end{array} \right),
\]
which, making use of Eq.\ (\ref{cons}), implies that
\begin{equation}
t' = t, \hspace{5ex} 0 = \frac{r'}{t} + \frac{r^{*}}{t^{*}}.
\label{symm}
\end{equation}
Hence, $r = r'$ if and only if $r/t$ is pure imaginary.

\section*{\tres 3. Rectangular barriers}
The reflection and transmission amplitudes for a given potential are
usually obtained by solving the time-independent Schr\"odinger
equation (\ref{sch}) (see, {\em e.g.}, Refs.\ 1--4). In the
exceptional case of a piecewise constant potential, the transfer
matrix (and, therefore, the reflection and transmission amplitudes)
can be readily obtained by means of matrix exponentiation.

The time-independent Schr\"odinger equation (\ref{sch}) can be
expressed as the first-order differential equation
\begin{equation}
\frac{{\rm d}}{{\rm d} x} \left( \begin{array}{c} \psi(x) \\
\psi'(x) \end{array} \right) = \left( \begin{array}{cc} 0 & 1 \\
v(x) - k^{2} & 0 \end{array} \right) \left( \begin{array}{c} \psi(x) \\
\psi'(x) \end{array} \right), \label{lin}
\end{equation}
where $v(x) \equiv 2m V(x)/\hbar^{2}$. Hence, if $V(x)$ is a
constant $V_{0}$ for $a \leq x \leq b$, the solution of Eq.\
(\ref{lin}) is
\[
\left( \begin{array}{c} \psi(x) \\ \psi'(x) \end{array} \right) =
\exp \left[ x \left( \begin{array}{cc} 0 & 1 \\ v_{0} - k^{2} & 0
\end{array} \right) \right] \left( \begin{array}{c} c_{1} \\ c_{2}
\end{array} \right),
\]
for $a \leq x \leq b$, where $v_{0} \equiv 2m V_{0}/\hbar^{2}$, and
$c_{1}$, $c_{2}$ are some constants. Thus,
\begin{equation}
\left( \begin{array}{c} \psi(a) \\
\psi'(a) \end{array} \right) = \exp \left[ - L \left(
\begin{array}{cc} 0 & 1 \\ v_{0} - k^{2} & 0 \end{array} \right)
\right] \left( \begin{array}{c} \psi(b) \\ \psi'(b) \end{array}
\right), \label{exp}
\end{equation}
with $L \equiv b-a$. Letting
\[
J \equiv \left( \begin{array}{cc} 0 & 1 \\ v_{0} - k^{2} & 0
\end{array} \right)
\]
one finds that $J^{2} = (v_{0} - k^{2}) I$, where $I$ is the unit $2
\times 2$ matrix, hence (see, {\em e.g.}, Ref.\ 7)
\begin{eqnarray}
\exp (- L J) \!\!\! & = & \!\!\! \left\{ \begin{array}{ll} \cosh (L
\sqrt{v_{0} - k^{2}}) \, I - \displaystyle \frac{\sinh (L
\sqrt{v_{0} - k^{2}})}{\sqrt{v_{0} - k^{2}}} \, J, & {\rm if\ }
v_{0} - k^{2} > 0, \\[2ex]
\cos (L \sqrt{k^{2} - v_{0}}) \, I - \displaystyle \frac{\sin (L
\sqrt{k^{2} - v_{0}})}{\sqrt{k^{2} - v_{0}}} \, J, & {\rm if\ }
v_{0} - k^{2} < 0, \\[2ex]
I - L J, & {\rm if\ } v_{0} - k^{2} = 0. \end{array} \right.
\label{sl}
\end{eqnarray}

On the other hand, from Eq.\ (\ref{sol}) we have
\[
\psi(a) = A_{1} + A_{2}, \hspace{3ex} \psi'(a) = {\rm i} k (A_{1} -
A_{2}), \hspace{3ex} \psi(b) = B_{1} + B_{2}, \hspace{3ex} \psi'(b)
= {\rm i} k (B_{1} - B_{2}),
\]
that is,
\begin{equation}
\left( \begin{array}{c} A_{1} \\ A_{2} \end{array} \right) =
\frac{1}{2} \left( \begin{array}{cc} 1 & - {\rm i}/k \\
1 & {\rm i}/k \end{array} \right) \left(
\begin{array}{c} \psi(a) \\ \psi'(a) \end{array} \right),
\hspace{5ex} \left(
\begin{array}{c} \psi(b) \\ \psi'(b) \end{array} \right) =
\left( \begin{array}{cc} 1 & 1 \\ {\rm i}k & -{\rm i}k
\end{array} \right) \left( \begin{array}{c} B_{1} \\ B_{2}
\end{array} \right). \label{chb}
\end{equation}
(Note that Eqs.\ (\ref{chb}) correspond to the continuity conditions
for $\psi$ and $\psi'$ at $x = a$ and $x = b$.)

Then, noting that
\[
\frac{1}{2} \left( \begin{array}{cc} 1 & - {\rm i}/k \\
1 & {\rm i}/k \end{array} \right) J \left( \begin{array}{cc} 1 & 1
\\ {\rm i}k & -{\rm i}k \end{array} \right) = \frac{1}{2 {\rm i} k}
\left( \begin{array}{cc} v_{0} - 2k^{2} & v_{0} \\ - v_{0} & - v_{0}
+ 2k^{2} \end{array} \right),
\]
from Eqs.\ (\ref{exp})--(\ref{chb}) one finds that, for a
rectangular potential barrier (or potential well)
\begin{equation}
V(x) = \left\{ \begin{array}{ll} 0, & {\rm if\ } x < a {\rm \ or\ }
x> b, \\ V_{0}, & {\rm if\ } a \leq x \leq b, \end{array} \right.
\label{pot}
\end{equation}
the transfer matrix is given by
\begin{equation}
M = \left\{ \begin{array}{ll} \cosh (L \sqrt{v_{0} - k^{2}}) \, I -
\displaystyle \frac{\sinh (L \sqrt{v_{0} - k^{2}})}{\sqrt{v_{0} -
k^{2}}} \frac{1}{2 {\rm i} k} \left( \begin{array}{cc} v_{0} -
2k^{2} & v_{0} \\ - v_{0} & - v_{0} + 2k^{2} \end{array} \right),
& {\rm if\ } v_{0} - k^{2} > 0, \\[3ex]
\cos (L \sqrt{k^{2} - v_{0}}) \, I - \displaystyle \frac{\sin (L
\sqrt{k^{2} - v_{0}})}{\sqrt{k^{2} - v_{0}}} \frac{1}{2 {\rm i} k}
\left( \begin{array}{cc} v_{0} - 2k^{2} & v_{0} \\ - v_{0} & - v_{0}
+ 2k^{2} \end{array} \right), & {\rm if\ } v_{0} - k^{2} < 0,
\\[3ex]
I - \displaystyle \frac{{\rm i} kL}{2} \left( \begin{array}{cc} 1 &
-1 \\ 1 & -1 \end{array} \right), & {\rm if\ } v_{0} - k^{2} = 0.
\end{array} \right. \label{trans}
\end{equation}
Note that owing to the definitions of the amplitudes $A_{1}$,
$A_{2}$, $B_{1}$, and $B_{2}$ in terms of the exponentials ${\rm
e}^{\pm {\rm i} k(x-a)}$ and ${\rm e}^{\pm {\rm i} k(x-b)}$, the
transfer matrices (\ref{trans}) depend on $a$ and $b$ only through
their difference $L = b-a$. The simplicity of the transfer matrices
(\ref{trans}) contrasts with the complexity of the expressions for
the reflection and transmission amplitudes obtained in the standard
manner (see, {\em e.g.}, Ref.\ 2, chap.\ 5). Note also that even
though we follow the conventions of Ref.\ 5, the transfer matrix
(\ref{trans}) does not agree with the amplitudes given in Eq.\ (12)
of Ref.\ 5.

It may also be noticed that, by allowing $\sqrt{v_{0} - k^{2}}$ to
become pure imaginary or taking the limit as $\sqrt{v_{0} - k^{2}}$
goes to zero, from the first expression in (\ref{trans}) one can
obtain the other two. Furthermore, one can verify directly that,
when $k$ is real, the transfer matrices (\ref{trans}) are of the
form $\left( \begin{array}{cc} \alpha & \beta \\ \beta^{*} &
\alpha^{*} \end{array} \right)$, with $|\alpha|^{2} - |\beta|^{2} =
1$ and therefore they indeed belong to SU(1,1). On the other hand,
Eqs.\ (\ref{lin})--(\ref{sl}) hold for $k$ real or complex and since
the trace of $J$ is equal to zero for any value of $k$ (even if
$V(x)$ was not real), the determinant of $\exp (-LJ)$ is equal to 1;
therefore, the transfer matrices (\ref{trans}) have determinant
equal to 1 also when $k$ is complex, though $M$ no longer belongs to
SU(1,1).

The example considered in this section also allows us to illustrate
the fact that making use of the transfer matrix one can find the
energies of the bound states or the quasinormal modes, by
considering pure imaginary or complex values of $k$, respectively.

In the case of the bound states of the potential well (\ref{pot})
with $V_{0} < 0$, we have $E < 0$ and writing $k = {\rm i} |k|$,
from Eq.\ (\ref{sol}) we see that in order for the wave function to
remain bounded, $B_{2} = 0$ and $A_{1} = 0$. Then Eq.\ (\ref{tm})
implies that $M_{11}$, the first entry of the diagonal of $M$, must
be equal to zero. Since the determinant of the transfer matrix is
equal to 1 (independent of the value of $k$), this last condition is
equivalent to saying that the off-diagonal entries of $M$ (which
have opposite signs) must be equal to $+1$ or $-1$. In the present
case $v_{0} - k^{2} < 0$, and from the second line of Eq.\
(\ref{trans}) we have
\[
\frac{\sin (L \sqrt{-|k|^{2} - v_{0}})}{\sqrt{-|k|^{2} - v_{0}}}
\frac{v_{0}}{2|k|} = \pm 1,
\]
which is equivalent to the conditions obtained in the textbooks
(see, {\em e.g.}, Refs.\ 1,2).

The so-called quasinormal modes correspond to complex values of $k$
for which there are no incident waves on the potential barrier but
only outgoing waves. In this case the solution of the
time-independent Schr\"{o}dinger equation are of the form
\begin{equation}
\psi(x) = \left\{ \begin{array}{ll} A_{2} {\rm e}^{- {\rm i}k(x-a)}, & x < a, \\
B_{1} {\rm e}^{{\rm i}k(x-b)}, & x > b, \\ u(x), & a \leq x \leq b, \\
\end{array} \right. \label{q}
\end{equation}
assuming that the real part of $k$ is positive ({\em cf.}\ Eq.\
(\ref{sol}), $A_{1}$ and $B_{2}$ are equal to zero so that there are
no ingoing waves on the barrier). Thus, as in the case of the bound
states, we have $M_{11} = 0$ and, making use of the first expression
in Eq.\ (\ref{trans}), we have
\[
\cosh (L \sqrt{v_{0} - k^{2}}) - \frac{\sinh (L \sqrt{v_{0} -
k^{2}})}{\sqrt{v_{0} - k^{2}}} \frac{v_{0} - 2k^{2}}{2 {\rm i} k} =
0
\]
which can also be expressed in the form
\begin{equation}
\cosh (L \sigma) - \frac{\sinh (L \sigma)}{\sigma} \frac{\sigma^{2}
- k^{2}}{2 {\rm i} k} = 0, \label{6}
\end{equation}
with the definition $\sigma \equiv \sqrt{v_{0} - k^{2}}$. Hence,
$\sigma^{2} + k^{2} = v_{0}$. Following Chandrasekhar [8], we
parameterize $k$ and $\sigma$ according to
\begin{equation}
k = Q \sin \alpha, \quad \sigma = Q \cos \alpha, \label{para}
\end{equation}
with $Q^{2} = v_{0}$ and $Q \geq 0$ (assuming $v_{0} \geq 0)$.
Substituting these expressions for $k$ and $\sigma$ into Eq.\
(\ref{6}) we have
\[
\cosh (L \sigma) - \sinh (L \sigma) \frac{Q^{2} \cos^{2} \alpha -
Q^{2} \sin^{2} \alpha}{2 {\rm i} Q^{2} \sin \alpha \cos \alpha} = 0,
\]
which is equivalent to
\begin{equation}
\cosh (L \sigma) + {\rm i} \sinh (L \sigma) \cot 2 \alpha = 0.
\label{8}
\end{equation}
Making use of the identities $\sin z = - {\rm i} \sinh ({\rm i} z)$,
$\cos z = \cosh ({\rm i} z)$, this last equation can be written as
$\sinh (L \sigma) \cosh ({\rm i} 2 \alpha) - \cosh (L \sigma) \sinh
({\rm i} 2 \alpha) = 0$, which is equivalent to
\[
\sinh (L \sigma - {\rm i} 2 \alpha) = 0
\]
and, therefore, $L \sigma - {\rm i} 2 \alpha = {\rm i} n \pi$, where
$n$ is an integer. Then, letting $\alpha = \alpha_{1} + {\rm i}
\alpha_{2}$ and $\sigma = \sigma_{1} + {\rm i} \sigma_{2}$, we have
\begin{equation}
L \sigma_{1} = -2 \alpha_{2}, \quad L \sigma_{2} = 2 \alpha_{1} - n
\pi. \label{11}
\end{equation}

From the relation $\sigma_{1} + {\rm i} \sigma_{2} = Q \cos
(\alpha_{1} + {\rm i} \alpha_{2})$ [see Eq.\ (\ref{para})] we obtain
\begin{equation} \label{12}
\sigma_{1} = Q \cos \alpha_{1} \cosh \alpha_{2}, \quad \sigma_{2} =
-Q \sin \alpha_{1} \sinh \alpha_{2}
\end{equation}
and, combining Eqs.\ (\ref{11}) and (\ref{12}), it follows that
\begin{equation}
-2 \alpha_{2} = L Q \cos \alpha_{1} \cosh \alpha_{2}, \quad 2
\alpha_{1} - n \pi = - L Q \sin \alpha_{1} \sinh \alpha_{2}.
\label{25}
\end{equation}
Hence,
\begin{equation}
\tan \alpha_{1} \tanh \alpha_{2} = \frac{2 \alpha_{1} - n \pi}{2
\alpha_{2}}. \label{26}
\end{equation}
Similarly, from Eq.\ (\ref{para}), we have $k_{1} + {\rm i} k_{2} =
Q \sin (\alpha_{1} + {\rm i} \alpha_{2})$, that is,
\[
k_{1} = Q \sin \alpha_{1} \cosh \alpha_{2}, \quad k_{2} = Q \cos
\alpha_{1} \sinh \alpha_{2}
\]
and, making use of Eqs.\ (\ref{25}),
\begin{equation}
k_{1} = - \frac{2 \alpha_{2} \tan \alpha_{1}}{L}, \quad k_{2} = -
\frac{2 \alpha_{2} \tanh \alpha_{2}}{L}. \label{16}
\end{equation}

By hypothesis, $k_{1} \geq 0$ and $Q \geq 0$, therefore from Eqs.\
(\ref{26}) and (\ref{16}) it follows that
\begin{eqnarray*}
{\rm if\ } \alpha_{2} > 0 \; \Rightarrow \; \tan \alpha_{1} \leq 0,
\quad 2 \alpha_{1} - n \pi \leq 0 & \Rightarrow & n > 0, \quad
\frac{\pi}{2} \leq \alpha_{1} \leq \pi, \\
{\rm if\ } \alpha_{2} < 0 \; \Rightarrow \; \tan \alpha_{1} \geq 0,
\quad 2 \alpha_{1} - n \pi \geq 0 & \Rightarrow & n \leq 0, \quad 0
\leq \alpha_{1} \leq \frac{\pi}{2}.
\end{eqnarray*}
Given a solution, $\alpha_{1}$, $\alpha_{2}$, of Eqs.\ (\ref{25}),
the values of $k_{1}$ and $k_{2}$ are determined by means of Eqs.\
(\ref{16}).

Since the time-independent Schr\"odinger equation (\ref{sch}) is
obtained assuming that the wave function has a time dependence of
the form $\exp (- {\rm i} Et/\hbar)$. When $k$ is complex, $E$ has a
negative imaginary part for $k_{1} > 0$ [$k_{2}$ is negative, see
Eq.\ (\ref{16})] that produces an exponential decay in time.

Denoting by $M^{(a,b)}$ the matrix appearing in Eq.\ (\ref{tm}), we
have the relation
\[
M^{(a,c)} = M^{(a,b)} M^{(b,c)},
\]
for any value of $c$. This relation together with Eq.\ (\ref{trans})
allow us to readily find the transfer matrix (or, equivalently, the
transmission and reflection amplitudes) for any piecewise constant
potential and from the condition $M_{11} = 0$, the bound states and
quasinormal modes can then be obtained, though the expressions will
be even more involved than the ones considered here.

\section*{\tres 4. Reflection and transmission of electromagnetic waves}
The behavior of a linearly polarized electromagnetic plane wave
normally incident on a slab of dielectric material can be found
following a procedure similar to that employed in the preceding
section. For plane monochromatic waves propagating along the
$x$-axis with the electric field parallel to the $y$-axis in a
homogeneous dielectric medium, the wave equation reduces to
\begin{equation}
\frac{{\rm d}^{2} E_{y}}{{\rm d}x^{2}} + k^{2} E_{y} = 0,
\label{em1}
\end{equation}
where $k = n \omega/c$, $n$ is the refractive index of the medium
and $\omega$ is the frequency of the wave. Equation (\ref{em1}) can
be expressed as the first-order equation
\begin{equation}
\frac{{\rm d}}{{\rm d} x} \left( \begin{array}{c} E_{y} \\ {\rm
d}E_{y}/ {\rm d}x \end{array} \right) = \left( \begin{array}{cc} 0 &
1 \\ - k^{2} & 0 \end{array} \right) \left( \begin{array}{c} E_{y}
\\ {\rm d}E_{y}/ {\rm d}x \end{array} \right), \label{em2}
\end{equation}
which is of the form (\ref{lin}) with $v = 0$; hence,
\[
\left( \begin{array}{c} E_{y}(a) \\ {\rm d}E_{y}/ {\rm d}x|_{x = a}
\end{array} \right) = \widetilde{M} \left( \begin{array}{c} E_{y}(b)
\\ {\rm d}E_{y}/{\rm d}x|_{x = b} \end{array} \right),
\]
where [see Eqs.\ (\ref{exp}) and (\ref{sl})]
\[
\widetilde{M} = \cos (kL)\, I - \frac{\sin (kL)}{k} \left(
\begin{array}{cc} 0 & 1 \\ - k^{2} & 0 \end{array} \right)
\]
and $L = b - a$.

If the slab is bounded by the planes $x = a$ and $x = b$ and, for
instance, surrounded by vacuum, Eq.\ (\ref{em1}) has solutions of
the form
\begin{equation}
E_{y} = \left\{ \begin{array}{ll} A_{1} {\rm e}^{{\rm i} k_{0}
(x-a)} + A_{2} {\rm e}^{-{\rm i} k_{0} (x-a)}, & {\rm for\ } x < a,
\\ B_{1} {\rm e}^{{\rm i} k_{0} (x-b)} + B_{2} {\rm e}^{-{\rm i}
k_{0} (x-b)}, & {\rm for\ } x > b, \end{array} \right. \label{em3}
\end{equation}
where $k_{0} \equiv \omega/c$. Faraday's law imply that the
$z$-component of the magnetic field is proportional to ${\rm
d}E_{y}/ {\rm d}x$ and, therefore, the continuity of the tangential
components of the fields at the boundary of the slab amounts to the
continuity of $E_{y}$ and ${\rm d}E_{y}/ {\rm d}x$ and from Eq.\
(\ref{em3}) we see that
\[
E_{y}(a) = A_{1} + A_{2}, \quad \frac{{\rm d}E_{y}}{{\rm d}x}(a) =
{\rm i} k_{0} (A_{1} - A_{2}), \quad E_{y}(b) = B_{1} + B_{2}, \quad
\frac{{\rm d}E_{y}}{{\rm d}x}(b) = {\rm i} k_{0} (B_{1} - B_{2}),
\]
thus, proceeding as in the previous section we obtain the relation
\begin{equation}
\left( \begin{array}{c} A_{1} \\ A_{2} \end{array} \right) = M
\left( \begin{array}{c} B_{1} \\ B_{2} \end{array} \right)
\label{tme}
\end{equation}
with the transfer matrix
\begin{equation}
M = \cos (kL) \, I - \frac{\sin (kL)}{k} \frac{{\rm i}}{2 k_{0}}
\left( \begin{array}{cc} k_{0}^{2} + k^{2} & - k_{0}^{2} + k^{2}
\\ k_{0}^{2} - k^{2} & - k_{0}^{2} - k^{2} \end{array} \right),
\label{em4}
\end{equation}
which is related to the transmission and reflection amplitudes as in
Eq.\ (\ref{su}). It may be noticed that, also in the present case,
the transfer matrix (\ref{em4}) belongs to SU(1,1) for $k$ real and
that the determinant of $M$ is equal to 1 even if $k$ is complex
(which would correspond to a nonzero conductivity).

\section*{\tres Acknowledgment}
One of the authors (I.R.G.) thanks the Vicerrector\'{\i}a de
Investigaci\'on y Estudios de Posgrado of the Universidad Aut\'onoma
de Puebla for financial support through the programme ``La ciencia
en tus manos.''

\section*{References}
\newcounter{ref} \begin{list}{\hspace{1.3ex}\arabic{ref}.\hfill}
{\usecounter{ref} \setlength{\leftmargin}{2em}
\setlength{\itemsep}{-.98ex}}
\item L.I. Schiff,  {\it Quantum Mechanics}, 3rd ed.\ (McGraw-Hill, New York, 1968).
\item S. Gasiorowicz, {\it Quantum Physics}, (Wiley, New York, 1974).
\item I.I.\ Gol'dman and V.D.\ Krivchenkov, {\it Problems in
Quantum Mechanics}, (Pergamon, London, 1961; reprinted by Dover, New
York, 1993).
\item K.\ Gottfried and T-M.\ Yan, {\it Quantum Mechanics:
Fundamentals}, 2nd ed., (Springer, New York, 2003).
\item L.L.\ S\'anchez-Soto, J.F.\ Cari\~nena, A.G.\ Barriuso and
J.J.\ Monz\'on, {\it Eur.\ J.\ Phys.}\ {\bf 26} (2005) 469.
\item J.R.\ Reitz, F.J.\ Milford and R.W.\ Christy, {\it Foundations of
Electromagnetic Theory}, 4th ed.\ (Addison-Wesley, Reading, MA,
1993).
\item K.B.\ Wolf, {\it Rev.\ Mex.\ F\'{\i}s.}\ {\bf 49} (2003) 465.
\item S. Chandrasekhar, {\it Proc.\ R.\ Soc.\ London} A {\bf 344} (1975) 441,
reprinted in {\it Selected Papers}, Vol.\ 6, (Chicago University
Press, Chicago, 1991).
\end{list}
\end{document}